\documentstyle[emulateapj]{article}
\newcommand{\gsim}{\lower .5ex\hbox{$\buildrel > \over {\sim}$}}
\newcommand{\lsim}{\lower .5ex\hbox{$\buildrel < \over {\sim}$}}

\def\wig#1{\mathrel{\hbox{\hbox to 0pt{%
          \lower.5ex\hbox{$\sim$}\hss}\raise.4ex\hbox{$#1$}}}}
\def\mj{M$_{\rm J}\ $}
\def\rj{R$_{\rm J}\ $}

\def\etal{{\it et~al.\,}}
\def\mo{M$_\odot$}
\def\ro{R$_\odot$}
\def\lo{L$_\odot\,$}
\def\mp{M$_{\rm p}$}
\def\rp{R$_{\rm p}\,$}

\def\mstar{M$_{\ast}$}
\def\teff{T$_{\rm eff}\,$}
\def\teffs{T$_{\rm eff}$s$\,$}
\def\Dwa{$\,$\uppercase\expandafter{\romannumeral5}$\,$}

\def\sles{\lower2pt\hbox{$\buildrel {\scriptstyle <}
   \over {\scriptstyle\sim}$}}
\def\sgreat{\lower2pt\hbox{$\buildrel {\scriptstyle >}
   \over {\scriptstyle\sim}$}}

\begin{document}

\slugcomment{Accepted to Ap.J. Letters}

\title{On the Radii of Close-in Giant Planets}

\author{A. Burrows\altaffilmark{1}, T. Guillot\altaffilmark{2}, W.B. Hubbard\altaffilmark{3}, 
        M.S. Marley\altaffilmark{4}, D. Saumon\altaffilmark{5}, J.I. Lunine\altaffilmark{3},
        and D. Sudarsky\altaffilmark{1}}

\altaffiltext{1}{Department of Astronomy, The University of Arizona,
                 Tucson, AZ \ 85721, USA (aburrows, sudarsky@as.arizona.edu).}
\altaffiltext{2}{Obs. de la C\^ote d'Azur, F-06304, Nice, Cedex 4, France (guillot@obs-nice.fr).}
\altaffiltext{3}{Lunar and Planetary Laboratory, The University of Arizona,
                 Tucson, AZ \ 85721, USA
                 (jlunine, hubbard@lpl.arizona.edu).}
\altaffiltext{4}{Department of Astronomy, New Mexico State University,
                 Box 30001/Dept. 4500, Las Cruces NM 88003, USA (mmarley@nmsu.edu).}
\altaffiltext{5}{Department of Physics and Astronomy, Vanderbilt University,
                 Nashville, TN 37235 USA (dsaumon@cactus.phy.vanderbilt.edu).}

\renewcommand{\thefootnote}{\fnsymbol{footnote}}
\setcounter{footnote}{1}

\begin{abstract}

The recent discovery that the close-in extrasolar giant planet, HD209458b, transits its 
star has provided a first-of-its-kind measurement of the planet's radius and mass. 
In addition, there is a provocative detection of the light reflected 
off of the giant planet, $\tau$ Boo b.  
Including the effects of stellar irradiation, we estimate the general behavior 
of radius/age trajectories for such planets
and interpret the large measured radii of HD209458b and $\tau$ Boo b in that context.
We find that HD209458b must be a hydrogen-rich gas giant.   Furthermore, the large radius of a close-in gas giant 
is not due to the thermal expansion of its atmosphere, but to the high residual entropy
that remains throughout its bulk by dint of its early proximity to a luminous primary.
The large stellar flux does not inflate the planet, but retards 
its otherwise inexorable contraction from a more extended configuration at birth.  
This implies either that such a planet was formed near its current orbital distance or that
it migrated in from larger distances ($\geq$0.5 A.U.), no later than a few times $10^7$ years of birth.

\end{abstract}


\section{Introduction}

The recent indirect detections of extrasolar giant planets (EGPs) by Doppler spectroscopy
have taught us that such planetary systems can be very much unlike our own.  To date, nearly $\sim$30
EGPs have been discovered around stars with spectral types from M4 to F7 
({\it e.g.}, \cite{mq95}; \cite{marcy99a}; \cite{fischer99}). 
The planets themselves have minimum masses (\mp $\sin (i)$) between $\sim$0.42 \mj and $\sim$10 \mj, orbital semi-major axes 
from $\sim$0.042 A.U. to $\sim$3.8 A.U., and eccentricities as high as $\sim$0.71.  Such variety
vastly expands the parameter space within which both theorists and observers must operate
in defining the physical character of extrasolar planetary systems.

The most interesting, unexpected, and problematic subclass of EGPs are those found within 
$\sim$0.1 A.U. of their primaries, 50--100 times closer than Jupiter is to our Sun.  At such orbital distances,
due to prodigious stellar irradiation alone, an EGP can have an effective temperature 
(\teff) greater than 600 K.  Indeed, the EGPs HD187123b, HD209458b,
$\tau$ Boo b, HD75289b, 51 Peg b, $\upsilon$ And b, and HD217107b likely all have \teffs above 1000 K.
This is to be compared with \teffs for Jupiter and Saturn of 125 K and 95 K, respectively.
The minimum masses of these close-in EGPs imply that they occupy the 
high-\teff/low-gravity corner of parameter space for which the compositions
and composition profiles may be unique (\cite{seager98}; \cite{gouk99}).  

However, high stellar fluxes on a close-in EGP can have profound structural consequences 
for the planet. In particular, stellar insolation can be responsible for maintaining the planet's radius
at a value 20\% to 80\% larger than that of Jupiter itself (\cite{guillot96}).  This prediction 
has recently been verified by the observation of the transit by HD209458b of its primary.  The depth of the transit yields
values for its radius that range from $\sim$1.27 \rj (\cite{char99b}) to $\sim$1.7 \rj (\cite{henry99}), with a best
value near $\sim$1.4 \rj (\cite{mazeh99}).  Importantly, since HD209458b transits its primary, astronomers can derive $\sin (i)$, 
from which the planet's mass can be directly determined.  This is a major advance in the emerging study of extrasolar planets.
In this paper, we focus on the HD209458 system and what broadly can be concluded theoretically 
from these new transit data.  We do not provide detailed models from which one can
extract bulk or atmospheric composition.  Rather, we show that the large 
radius of HD209458b is a consequence of the retardation
by stellar irradiation of the otherwise natural cooling of the convective core of the planet
and that such a radius requires that the planet did not dwell for long, if at all,
at large orbital distances after its formation.  Hence, the large radius of a close-in EGP
is not due to the thermal expansion of its atmosphere, but to the high residual entropy
that remains throughout its bulk as a consequence of its early proximity to a luminous primary.

Recently, Cameron \etal (1999), using spectral deconvolution, claim to have seen $\tau$ Boo b in reflection.
To investigate this exciting possibility, we calculate a range of theoretical radii for the $\tau$ Bo\"otis planet.
As we show, the Cameron \etal value of 1.6--1.8 \rj for $\tau$ Boo b's radius, if verified,
is a challenge to the still embryonic theory of massive, close-in EGPs.  
Nevertheless, with the discovery of a transiting planet, HD209458b,  
with the maturation of the technique of spectral deconvolution,
and with the anticipated development of adaptive optics and interferometry for the direct 
study of extrasolar planets (\cite{angel94}), we are clearly entering a new phase in extrasolar
planetary research.

\section{A Summary of the Data \label{data}}

The transit of the F8V/G0V star HD209458 (at a distance of 47 parsecs) by HD209458b lasts 
$\sim$3 hours (out of a total period of 3.524 days) and has a depth 
of $\sim$1.5-2.0\%.  Its ingress and egress phases each last  
$\sim$25 minutes (\cite{char99b}; \cite{henry99}). The properties of 
the planet, in particular its orbital distance and its radius,  
scale with the properties of the star and the most recent study of HD209458 was conducted
in the context of these transits by Mazeh \etal (2000).  They conclude from $log {\rm g}$/\teff
spectral-line fits and $M_{V}$/$(B-V)$ photometric fits 
that \mstar = $1.1\pm0.1$ \mo,  R$_{\ast}$ = $1.2\pm 0.1$ \ro, \teff$\sim$6000 K,
[Fe/H]$\sim$0.0, $t = 5.5\pm 1.5$ Gyr, and L$_{\ast} \sim 2.0$ \lo.  Mazeh \etal then derive for the planet:
\rp = 1.40$\pm0.17$ \rj, \mp = $0.69\pm0.05$ \mj ($\propto$M$_{\ast}^{2/3}$), 
$i = 86^{\circ}.1\pm1^{\circ}.6$, and $a=0.047$ A.U. ($\propto$M$_{\ast}^{1/3}$).
(All symbols have their standard meanings.)

Fuhrmann \etal (1998) provide parameters for the F7V star $\tau$ Bo\"otis:   
\mstar = $1.42\pm0.05$ \mo, R$_{\ast}$ = $1.48\pm 0.05$ \ro, \teff$\sim$6360 K,
[Fe/H]$\sim$+0.27$\pm$0.08, $t = 1.0\pm 0.6$ Gyr, and L$_{\ast} \sim 3.2$ \lo.
Its Hipparcos distance is $\sim$15.6 parsecs.
With the Fuhrmann \etal mass for $\tau$ Bo\"otis, Butler \etal (1997) would have obtained 
for its close-in EGP: \mp$\sin (i)$ = 4.33 \mj, $a\sim 0.049$ A.U., and $P = 3.313$ days.
Charbonneau \etal (1999a) quote an upper limit at $\sim$4900\AA\ of $5\times10^{-5}$ to 
the fraction of the star's light reflected off the planet.
Cameron \etal (1999) claim to have detected in the blue-green region 
of the spectrum a reflected fraction of $1.9\pm0.4\times10^{-4}$.
From the semi-amplitude ($\sim$74 km s$^{-1}$) of 
the Doppler shift of this reflected fraction, 
they derive an orbital inclination ($i$)
of 29$^{\circ}$, which would yield a mass for $\tau$ Boo b of $\sim$9 \mj.
From their reflected fraction, an orbital planetary phase function,
and a geometric albedo ($A_g$) of 0.55 (similar to that of Jupiter in the visible), Cameron \etal obtain
a radius for $\tau$ Boo b of 1.6--1.8 \rj.  If $A_g$ were smaller, the 
inferred radius of the planet would be larger ($\propto A_g^{-1/2}$).

\section{The Radii of Close-in EGPs}

By whatever processes giant planets are initially assembled, 
they must start out significantly larger than they end up.  In isolation,
they would cool inexorably due to radiation from their surfaces and shrink 
accordingly, just as does a protostar or a brown dwarf.  Early on,
due to the negative effective specific heat of an object in hydrostatic 
equilibrium supported by ideal gas pressure, energy loss results in an increase in its central
temperature.  However, the density increases more quickly and, as a consequence, the specific entropy ($S$)
monotonically decreases.  Since EGPs are almost fully convective (even if under
significant stellar irradiation; \cite{guillot96}), they are isentropes.  
This is an essential point.  Given an EOS and a planetary mass, an EGP's
core entropy determines its radius (and its surface gravity).  
A large radius is a consequence of a large entropy. 

As Zapolsky and Salpeter (1969) demonstrated for planets made 
of high-$Z$ material, any planet with a cold radius larger
than $\sim$0.5 \rj, must be made predominantly of hydrogen.  Using the ANEOS equation of state tables (\cite{t90}), 
we derive that an ``olivine" (rock) or H$_2$O (ice) planet with a mass of 0.69 \mj has a radius 
of 0.31 \rj or 0.45 \rj, respectively.  Importantly, these radii are 3--4 times smaller than observed for HD209458b
and prove that HD209458b must be a hydrogen-rich gas giant; it cannot be a giant terrestrial planet or 
an ice giant such as Neptune or Uranus. 

The rate with which an EGP shrinks is determined by the opacities in its
outer radiative zone and the degree of stellar irradiation.  
If there were no radiative losses, the EGP would not shrink.  In isolation,
the energy loss rate and \teff are determined in the context of a self-consistent
radiative/convective model and \teff is a function of only $S$ and \mp. 
(Note that, for a given metallicity, the surface gravity of the EGP is a 
function of $S$ and \mp\ alone.) The atmospheric flux in the 
outer skin determines the temperature profile at the boundary of the 
convective core and, hence, the rate of core entropy and radius decrease.
According to theory (\cite{saumon96}; Burrows \etal 1995,1997), 
isolated EGPs shrink rapidly.  A 1-\mj EGP in isolation
contracts below 2.0 \rj in less than $10^6$ years.  The theory depends
upon the atmospheric opacities and metallicity and can reproduce the current Jupiter (\cite{hub99}),
but there still remain major uncertainties concerning grain and cloud formation (\cite{lunine89}), rainout (\cite{burrows99}), 
gas-phase opacities (\cite{burrows97}), and the depth of 
penetration of the stellar flux (\cite{guillot00}). Hence, while 
the basic theory of EGP evolution is firm, the details are not and significant ambiguities 
in the variation of \rp and \teff with age persist. 

It is by altering the temperature/pressure profile of the atmosphere of
an EGP that stellar irradiation can retard the evolution of the EGP's core entropy and, hence,
\rp.  Essentially, irradiation flattens the temperature profile at
the top of the convective zone, while at the same time moving the radiative/convective boundary inward.
The consequent growth of the radiative zone is a central feature of the large-radius phenomenon (\cite{guillot96}). 
Since radiative fluxes are governed by the product of thermal diffusivities and temperature
gradients and since the thermal diffusivity decreases with increasing pressure, 
the flux of energy out of the convective core and the rate of core entropy change are 
reduced.  For close-in EGPs, \teff stabilizes early at large values,
while the ``effective" \teff of the core, where most of the heat and mass resides,
is drastically lowered.  The upshot is a retardation of the contraction of the
planet.  Large EGP radii are a consequence of such retardation, and not of the expansion
of the outer envelope by stellar heating.  This is easy to demonstrate by noting
that the scale height of HD209458b's atmosphere, under the assumption that \teff 
is between 1200 K and 1700 K, is only $\sim$1\% of \rp.
Though the mapping between $S$ and \rp is unaltered by irradiation, the mapping
between $S$ and age can change significantly.  Guillot \etal (1996) predicted
this behavior for 51 Peg b, using a Bond albedo ($A_B$) of 0.35, obtaining \rp{'s} of 1.1--1.3 \rj
after 8 Gyr.  Scaling the results of that paper using the stellar flux on HD209458b
and a mass of 0.69 \mj, we obtain \rp{'s} between 1.4 \rj and 1.6 \rj for ages between
10 and 3 Gyr, perfectly in line with the transit observations (\S 2).
Figure 1 depicts two possible evolutionary trajectories for HD209458b, if born and fixed at 0.049 A.U.
Also included on Figure 1 is a theoretical \rp{-$t$} trajectory for a 0.69-\mj EGP in isolation.
We have used for these calculations the formalisms of Guillot \etal (1995, 1996), Guillot and Morel (1995), 
and Burrows \etal (1997).
Tidal and Roche effects have been ignored and are relevant only at very early ages ($\wig{<}$$10^4$ yrs).
Note, however, that despite its small orbital distance, HD209458b is beyond the Roche limit by a factor of $\sim$2.5
and is stable against loss both by thermal (Jeans)
escape and non-thermal processes involving absorption of stellar UV flux (\cite{trilling98}).
A box of empirical ages and \rp{'s} (Mazeh \etal 2000) is superposed.

As can be seen in Figure 1, there is a great difference between the theoretical radius  
of an isolated and an irradiated EGP.  Importantly, since the scale-height effect is miniscule,
the \rp{-$t$} trajectory of the isolated EGP (model I) immediately suggests that if HD209458b were
allowed to dwell at large orbital distances ($\geq$0.5 A.U.) for more than a few
$\times 10^7$ years, its observed radius could not be reproduced.  As Figure 1 demonstrates, 
it is at such ages that the radius of an isolated 0.69-\mj EGP falls below
HD209458b's observed radius.  Note that for $\geq$0.5 A.U. the intrinsic
luminosity of our isolated HD209458b model at such an age exceeds the amount of 
stellar light that would have been intercepted.  Hence, stellar irradiation would have
had little effect on this model.  However, many orbital-distance/age histories can be 
contemplated and these will be the subject of a future work (\cite{guillot00}).
To reiterate, HD209458b could not have cooled off and achieved a radius or an entropy
like that of Jupiter and then moved in. If it migrated, it had to have moved in early
in its life.  The radius of a close-in EGP depends upon its history (and the history of the star); large 
radii require early proximity to a central star.  This fact provides an upper limit to 
the timescale of planetary migration, if migration did indeed occur (\cite{trilling}):
conservatively, HD209458b dwelled less than a few$\times 10^7$ years at more than 0.5 A.U. 

The actual evolution of a close-in EGP's radius depends upon its Bond albedo (\cite{marley99};
\cite{seager98}; \cite{sud99}), the level of any clouds formed and their optical depths, 
the gas-phase abundances and opacities, the helium and metallicity fractions, 
the H-He EOS (\cite{scvh95}; \cite{saumon99}), the depth of stellar flux penetration, and the primary star.
In a later paper (\cite{guillot00}), we explore these effects and conduct a detailed parameter study.
A range of trajectories similar to those depicted in Figure 1 are still possible
and when data on the HD209458b transit and primary star are further refined, theorists
may well be able to sharply constrain the character of HD209458b's atmosphere and composition.

\subsection{Theoretical Radius for $\tau$ Boo b}

We provide in Figure 1 two representative theoretical \rp{-$t$} trajectories for $\tau$ Boo b, along
with an error box constructed using Cameron \etal (1999) and Fuhrmann \etal (1998).  Depending
upon L$_{\ast}$ and its Bond albedo, which might vary from $\sim$0.0 to perhaps 0.6, $\tau$ Boo b's \teff is between 1350 K and 1750 K.
These \teffs are higher than those corresponding to HD209458b and reflect $\tau$ Bo\"otis' higher L$_{\ast}$.
Due to uncertainties in the relative position of the silicate and iron cloud decks and the region of
neutral alkali metals, there are still uncertainties in the atmospheric
composition and albedos of such an EGP (\cite{marley99}; \cite{sud99}).  
A high Bond albedo might be a consequence of
the presence at altitude of reflecting clouds.  Without these clouds, both the Bond albedo and the
geometric albedo in the blue-green region of the planetary spectrum
would be low, perhaps below 0.1.  A low geometric albedo would put the measured
radius of $\tau$ Boo b even higher than the range quoted by Cameron \etal.  As the evolutionary
models on Figure 1 suggest, though a lower Bond albedo results in a larger \rp at a given age,  
the \rp{-$t$} error box for $\tau$ Boo b given in Figure 1 still seems out of reach.
Note that a 9-\mj $\tau$ Boo b model in isolation achieves a radius of 1.6 \rj within a scant 5 Myr
and that at 1 Gyr such a model has a radius of 1.1 \rj.  
 
The early radii of $\tau$ Boo b shown on Figure 1 are so much smaller than 
those of HD209458b because of the larger inferred mass of $\tau$ Boo b and 
the particular choice of opacity data used for these exploratory models.
Other opacity databases would give quantitatively different radius-age 
trajectories early on, but would not alter the conclusion that the theoretical
radius of a massive planet at late times ($>$ few$\times 10^7$ years) is significantly below
the $\tau$ Boo b error box shown in Figure 1.  Furthermore, under the assumptions that
the Bond albedo is zero, that there is no outer radiative zone, and that 
the atmosphere and core of the planet have the same entropy, we can obtain a strong
upper bound to the radius of a planet of a given age and mass (Guillot \etal 1996).   
For $\tau$ Boo b, this radius is 1.58 \rj for 7 \mj and 1.48 \rj for 10 \mj,
both below the quoted radius range.
Hence, we have difficulty fitting the Cameron \etal (1999) reflection data.  
In particular, despite significant irradiation, the large planetary mass measured by Cameron \etal
(7-10 \mj) results in rapid early contraction.
If the Bond albedo is lower, the theoretical $\tau$ Boo b radius for an age near 1 Gyr 
would be larger, but the measured radius would also be larger. 

\section{The Effects of the Equation of State}

Our cooling calculations use the hydrogen/helium EOS of Saumon, Chabrier, and Van Horn (1995, SCVH).
Recent shock-compression experiments on deuterium (\cite{holmes95}; \cite{silva97}; \cite{collins98})
show that this EOS underestimates the degree of molecular dissociation
for pressures in the range $0.1 \wig< P \wig< 2\,$Mbar.  However, by softening the repulsive part of the
H$_2$--H and H--H potentials, the EOS of SCVH can be made to reproduce all
experimental results (\cite{saumon99}).
Since adiabats of hot Jupiters computed with the modified EOS are systematically
cooler and denser than those of SCVH by up to 6\% in $T$ and 10\% in $\rho$,
such changes in the EOS will reduce the theoretical radius of HD209458b by only a few percent,
while modifying the corresponding quantity for $\tau$ Boo b by a negligible amount.

\newpage

\section{Atmospheric Transmission and Refraction}

Details of the transit lightcurve depend upon the
distribution of slant optical depth, as determined by molecular
opacity and cloud layers, and, to a lesser extent,
on refractive redistribution of the stellar surface brightness
by the planetary atmosphere.

We use a theory to compute refractive effects that is essentially
identical to that of the standard theory for occultations of
stars by planetary atmospheres (\cite{hub90})
and have calculated the stellar brightness distribution 
for the transit of a planet with a radius equal to 1.4
R$_J$ and a hot solar-composition atmosphere.  
The slant optical depth $\tau$
is computed for molecular opacity sources (with and without clouds) and at a
variety of wavelengths.  
The effective radius of the planet, as determined
by fitting the transit lightcurve, will depend upon the
gas-phase opacity, the molecular composition, and the location of the optically-thick cloud
layers (\cite{seager99a}).  It will also be a diagnostic function of wavelength.
Our preliminary calculations for cloud-free models indicate that $|\Delta R_{\rm p}/R_{\rm p}|$ 
between 4500 \AA\ and 6500 \AA\ might be as much as 3\%, smaller ($\sim$0.5\%) if the alkali metals
are important.  
Using a simple model for silicate cloud growth (\cite{lunine89}; \cite{marley99}),
we find a cloud base for HD209458b between $10^{-2}$ and 0.4 bars
and grain particle sizes between 1 and 100 microns.  While for cloud models the 
wavelength dependence is muted, since 
the opacity varies strongly with particle size, there exists
the possibility of remote sensing of cloud properties with high-precision measurements
of upcoming transits (\cite{hubbard20}).

\section{Conclusions}

Our theoretical calculations 
allow us to draw several specific conclusions:

\begin{enumerate}

\item HD209458b is a real object, made predominantly of hydrogen.

\item HD209458b's radius is a consequence of the retardation of contraction by stellar irradiation
      and is not due to atmospheric expansion by stellar heating.

\item A large radius such as that of HD209458b requires early proximity to its central star.

\end{enumerate}

Curiously, given L$_{\ast}$, $a$, and \rp, HD209458b's total luminosity is 
$\sim 1.5\times 10^{-4}$ \lo, about twice that of a star with $\sim$100 times
the mass at the very bottom of the stellar main sequence. 
Given the large inferred mass of $\tau$ Boo b, its large radius is less easy to explain theoretically. 
However, the inherent difficulties of close-in EGP modeling 
may yet be responsible for theoretical surprises of a qualitative nature.

\acknowledgements

We thank Tim Brown, David Charbonneau, Geoff Marcy, 
Michel Mayor, Sara Seager, Richard Freedman and Jim Liebert for many useful discussions
and for the use of data in advance of publication. This work was supported in part by 
NASA grants NAG5-7499, NAG5-7073, NAG5-4214, NAG5-7211, NAG2-6007, 
NAG5-4988, and NAG5-4987, as well as by an NSF CAREER grant (AST-9624878) to M.S.M.

\clearpage

\begin{figure}
\epsscale{1.00}
\plotone{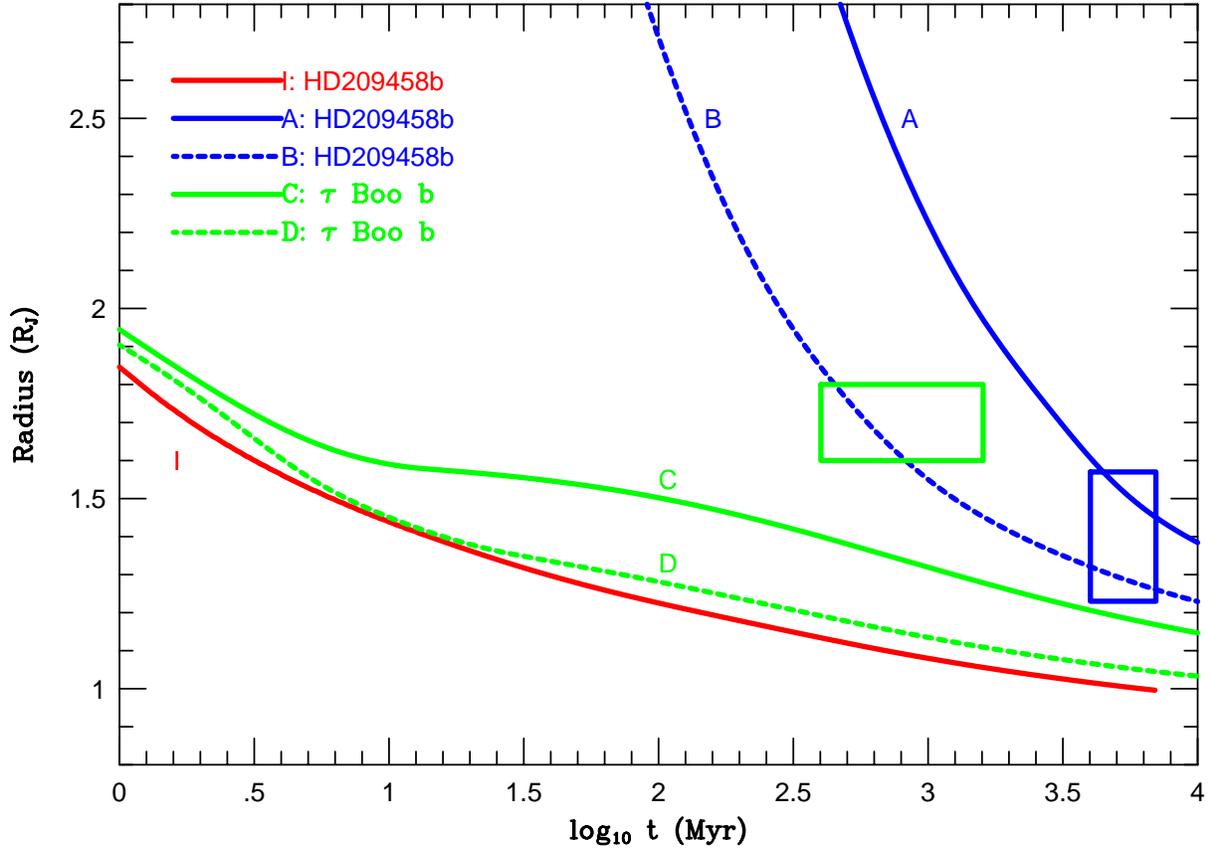}
\caption{
Theoretical evolution of the radii of HD209458b and $\tau$ Boo b (in \rj; $\sim 7\times 10^4$ km) with age (in Gyrs).
Model I (solid) is for a 0.69 \mj object in isolation (Burrows \etal 1997).  Models A ($A_B = 0.0$;
\teff $\sim$1600 K) and B ($A_B = 0.5$; \teff $\sim$1200 K)) are for a close-in, irradiated HD209458b
at its current orbital distance from birth, using the opacities of Alexander and Ferguson (1994).
Models C ($A_B = 0.0$; \teff$\sim$1750 K; \mp = 7 \mj) and D ($A_B = 0.5$; \teff$\sim$1350 K; \mp = 10 \mj)
are for a close-in, irradiated $\tau$ Boo b, using a similar opacity set.  The formalism
of Guillot \etal (1996) was employed for models A-D.  The ranges spanned by models A and B and by models C and D
for HD209458b and $\tau$ Boo b, respectively, reflect the current ambiguities in the observations
and in the theoretical predictions due to cloud, opacity, composition, and depth of flux penetration uncertainties.
Superposed are an error box for HD209458b (far right) using the data of Mazeh \etal (2000) and one for
$\tau$ Boo b using the data of Cameron \etal (1999) and Fuhrmann \etal (1998).
}
\end{figure}

\end{document}